\documentclass[15pt,preprint]{aastex}

\usepackage{graphicx,epsfig,natbib,color}

\usepackage{graphicx}
\usepackage{epsfig}
\usepackage{natbib}

\begin{document}
\title{Re-flaring of a Post-Flare Loop System Driven by Flux Rope Emergence and Twisting}

\author{X. Cheng\altaffilmark{1,2}, M. D. Ding\altaffilmark{1,2},
Y. Guo\altaffilmark{1,2}, J. Zhang\altaffilmark{1,3}, J.
Jing\altaffilmark{4}, and T. Wiegelmann\altaffilmark{5}}

\affil{$^1$ Department of Astronomy, Nanjing University, Nanjing
210093, China} \email{dmd@nju.edu.cn}

\affil{$^2$ Key Laboratory of Modern Astronomy and Astrophysics
(Nanjing University), Ministry of Education, Nanjing 210093, China}

\affil{$^3$ Department of Computational and Data Sciences, George
Mason University, 4400 University Drive, MSN 6A2, Fairfax, VA 22030,
USA}

\affil{$^4$ Space Weather Research Laboratory, New Jersey Institute
of Technology, Newark, NJ 07102, USA}

\affil{$^5$ Max-Planck-Institut f\"ur Sonnensystemforschung,
Max-Planck-Strasse 2, 37191 Katlenburg-Lindau, Germany}

\begin{abstract}

In this letter, we study in detail the evolution of the post-flare
loops on 2005 January 15 that occurred between two consecutive solar
eruption events, both of which generated a fast halo CME and a major
flare. The post-flare loop system, formed after the first CME/flare
eruption, evolved rapidly, as manifested by the unusual accelerating
rise motion of the loops. Through nonlinear force-free field (NLFFF)
models, we obtain the magnetic structure over the active region. It
clearly shows that the flux rope below the loops also kept rising
accompanied with increasing twist and length. Finally, the
post-flare magnetic configuration evolved to a state that resulted
in the second CME/flare eruption. This is an event in which the
post-flare loops can re-flare in a short period of $\sim$16 hr
following the first CME/flare eruption. The observed re-flaring at
the same location is likely driven by the rapid evolution of the
flux rope caused by the magnetic flux emergence and the rotation of
the sunspot. This observation provides valuable information on
CME/flare models and their prediction.

\end{abstract}

\section{Introduction}
The developed systems of loops, named post-flare loops, usually
occur in the decay phase of long duration events
\citep[e.g.,][]{bruzek64,kahler77,Harra98}. It is generally believed
that magnetic reconnection plays an important role in the formation
and evolution of the post-flare loops. The rising of post-flare
loops in the corona, as well as the separation of flare ribbons at
the footpoints, can be well explained by the classical reconnection
model, i.e., the CSHKP model
\citep{carmichael64,sturrock66,hirayama74,kopp76}. In this model,
the observed motions are caused by the systematic ascending of the
reconnection site in the corona \citep[e.g.,][]{cheng10a}. The
post-flare loops comprise of a system of loops heated and formed
consecutively through the ongoing reconnection process. Based on the
CSHKP model, the post-flare loops rise with a gradually decreasing
speed. However, \citet{svestka96} found some exceptional cases, in
which the post-flare loops rose with a constant speed  for a long
period of time, and was explained through the combination of two
entirely different processes: the initial development of the
magnetic reconnection proposed by \citet{kopp76} and the subsequent
expanding motion into the corona. It is generally believed that the
post-flare loops evolve without further energy input after their
formation. Therefore, they usually fade away into the background
after a certain time. To our knowledge, there is no report on the
re-activation of a post-flare loop system in previous studies.

Coronal mass ejections (CMEs) are another kind of large-scale solar
activity that may involve the magnetic field lines that twisted each
other, named flux rope, in their core, as revealed by many
coronagraph observations \citep[e.g.,][]{pick06}. The role of flux
ropes in the CME eruptions has also been studied extensively through
numerical simulation. \citet{amari00} simulated the evolution of a
flux rope and found that it plays a crucial role in the process of
the CME eruption. The flux rope was also used by \citet{fan01} and
\citet{fan03,fan04} to investigate the dynamic evolution of the
coronal magnetic field in response to the emergence of twisted
magnetic structures. Moreover, \citet{torok03,torok05} and
\citet{Kliem06} studied the instability of the flux rope and
proposed that the CME eruption can be initiated through the kink
and/or torus instability. \citet{cheng10b} suggested that the flux
rope may provide a magnetic structure favorable for the eruption and
regarded the formation period of the flux rope prior to the eruption
as the build-up phase of CMEs \citep[see also][]{su09,Aulanier10}.
\citet{schrijver08} and \citet{schrijver09} showed that the
emergence of current-carrying flux ropes can lead to the occurrence
of a series of major flares within a few hours in the same active
region. Further, \citet{canou09} constructed a flux rope structure
above magnetic polarity inversion line (PIL) of an emerging sunspot
by a nonlinear force-free field (NLFFF) model using the THEMIS
vector magnetogram as the input. Recently, \citet{guo10} also
obtained a flux rope structure through NLFFF modeling and found that
the flux rope coincided with the active region filament. However, up
to now, there is still no study on the time evolution of
extrapolated flux ropes.

In this letter, we report the re-flaring of a post-flare loop system
occurred on 2005 January 15, and further study the temporal
development of associated flux rope. In section 2, we present the
observations and data analysis method. Our results are shown in
section 3 followed by summary and discussions in section 4.

\begin{figure*}[!ht]
\begin{centering}
\epsscale{0.9} \plotone{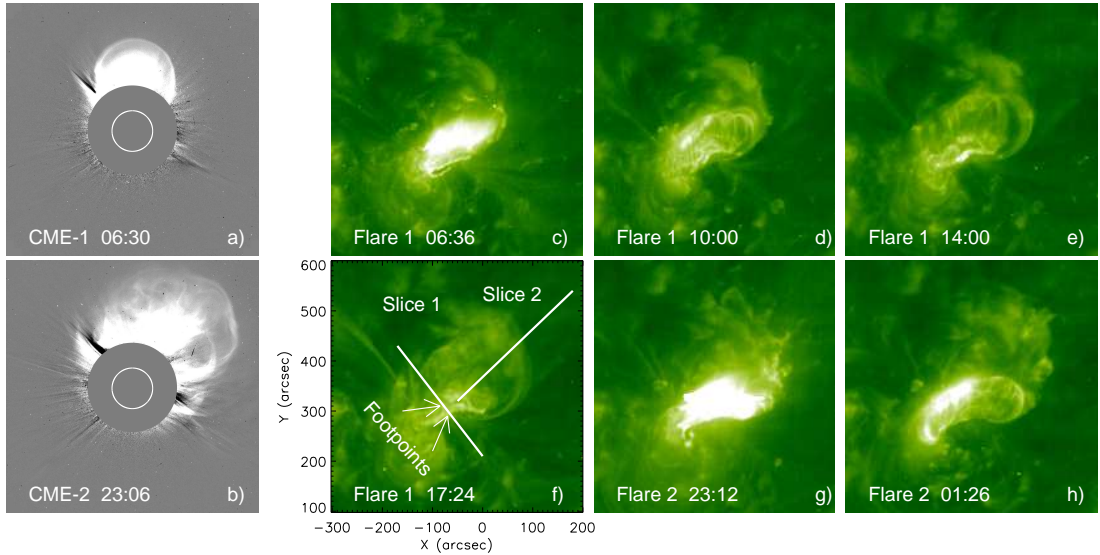} \caption{(a--b) Running
difference images of CMEs 1 and 2 observed by LASCO. (c) EIT 195
{\AA} image of the CME-associated Flare 1. (d--f) Three images
showing the post-flare loops after the peak time of flare 1. Slices
1 and 2 in panel (f) are used to trace the time evolution of the
post-flare loop system. (g) EIT 195 {\AA} image of the
CME-associated flare 2. (h) Image showing the post-flare loops of
flare 2. \label{cmeflare}}

(Animation of the post-flare loops is available in the online
journal)
\end{centering}
\end{figure*}

\section{Observations and Data Analysis}
On 2005 January 15, there were two halo CMEs (CME 1 and CME 2) that
were observed by the C2 and C3 coronagraphs of the Large Angle and
Spectrometric Coronagraph (LASCO) \citep{Brueckner95} on board the
$Solar$ $and$ $Heliospheric$ $Observatory$ ($SOHO$). They first
appeared in the C2 field of view (FOV) at 06:30 UT and 23:06 UT,
respectively, as shown in Figure 1a and 1b. Their linear fitting
speeds\footnote{http://cdaw.gsfc.nasa.gov/CME\_list} in the FOV of
LASCO both exceeded 2000 km s$^{-1}$. After visually inspecting the
Extreme-ultraviolet Imaging Telescope (EIT; Delaboudini\`{e}re et
al. 1995) 195 {\AA} data, we find that the two extremely fast CMEs
were associated with two major long-decay flares (Flare 1 and Flare
2), respectively, as seen in Figure 1c and 1g. The most important
finding is that the post-flare loop system of flare 1 was always
expanding and rising during the period between the two flares until
part of it was activated as the second CME/flare eruption, as
denoted in Figure 1c--g. The overall properties of the two CMEs and
associated flares are summarized in Table 1. The kinematics of the
post-flare loop system will be studied in the next section.

\begin{table*}
\caption{Properties of the two CME/flare events on 2005 January 15.}
\label{tb}
\begin{tabular}{ccccccccc}
\\ \tableline \tableline

CME &Time ${}^\mathrm{a}$
&Speed &Width &Flare &Onset &Peak &End &Magnitude  \\
&(UT) &(km s$^{-1}$)& (deg)& &(UT) &(UT) &(UT) &  \\

\tableline
CME 1 &06:30 &2049 &360  &Flare 1 &05:54 &06:38 &07:17 &M8.6\\
CME 2 &23:06 &2861 &360  &Flare 2 &22:25 &23:02 &23:31 &X2.6\\
\tableline
\multicolumn{9}{p{12.5cm}}{${}^\mathrm{a}$ Time of the first appearance in C2.}\\
\end{tabular}
\end{table*}

Inspecting the magnetic field data, we find that the two
flare-associated CMEs originated from the active region NOAA 10720,
which appeared on 2005 January 10 and then grew rapidly before it
was nearly at the disk center on 2005 January 10, and finally
decayed on January 20. The Michelson Doppler Imager (MDI; Scherrer
et al. 1995) magnetograms with 96 min cadence provides the
line-of-sight magnetic field evolution of this active region.
Fortunately, vector magnetic field data with a higher cadence of
$\sim$1 min and a higher spatial resolution of
$\sim$0.6\arcsec\,were obtained by the Digital Vector Magnetograph
(DVMG) at BBSO on January 15, which covered a FOV of $\sim
300\arcsec \times 300\arcsec$. The sensitivity of the line-of-sight
and transverse magnetic field is 2 and 20 G, respectively. Due to
the filter observation, the data are saturated when the field
strength exceeds $\sim 1000$ G. We fill these saturated areas with
the co-aligned MDI line-of-sight magnetic field to reduce its
influence on the extrapolation. The projection effect is negligible
since the vector magnetic field was observed at the disk center.
More details about the DVMG data can be found in \citet{jing09}.
Finally, we resolve the 180$^{\circ}$ azimuthal ambiguity of the
transverse magnetic field using the minimum energy algorithm
\citep{metcalf94,metcalf06}, which simultaneously minimizes both the
electric current density and the magnetic field divergence. This
state-of-art automatic algorithm gives a satisfactory resolution of
the 180$^{\circ}$ ambiguity, where the changes of the field
directions are smooth and the field divergence is sufficiently small
in the FOV of the DVMG.

Based on the DVMG data, we reconstruct the three-dimensional (3D)
coronal magnetic field by the NLFFF model using the optimization
algorithm; the model was originally proposed by \citet{wheatland00}
and implemented by \citet{Wie04}. To deal with the inconsistency
between the forced photospheric magnetic field and the force-free
assumption of the NLFFF models, as well as the noises of the
observed magnetic field, a preprocessing procedure was proposed by
\citet{Wie06}. The preprocessing procedure minimizes the net force
and torque of the photospheric magnetic field and keeps the
consistency between the final preprocessed data and the measured
data, given that the magnetic flux is balanced in the FOV. However,
the flux balance condition is often not satisfied because of the
observational limitation. \citet{guo10} showed that applying the
preprocessing procedure on the magnetic field in the original FOV is
better than embedding it in a larger FOV where the transverse
components of the magnetic field are unknown. Experiments of
embedding vector magnetograms into line-of-sight magnetograms have
been done in \citet{derosa09}. It was found that the missing
transverse magnetic field information usually lead to inconsistent
boundary conditions. Finally, the preprocessed data are submitted to
the optimization method as the bottom boundary to extrapolate the
NLFFF.

\section{Expansion of Post-flare Loops and Magnetic Field Evolution}

\begin{figure*}[!ht]
\begin{centering}
\epsscale{0.48} \plotone{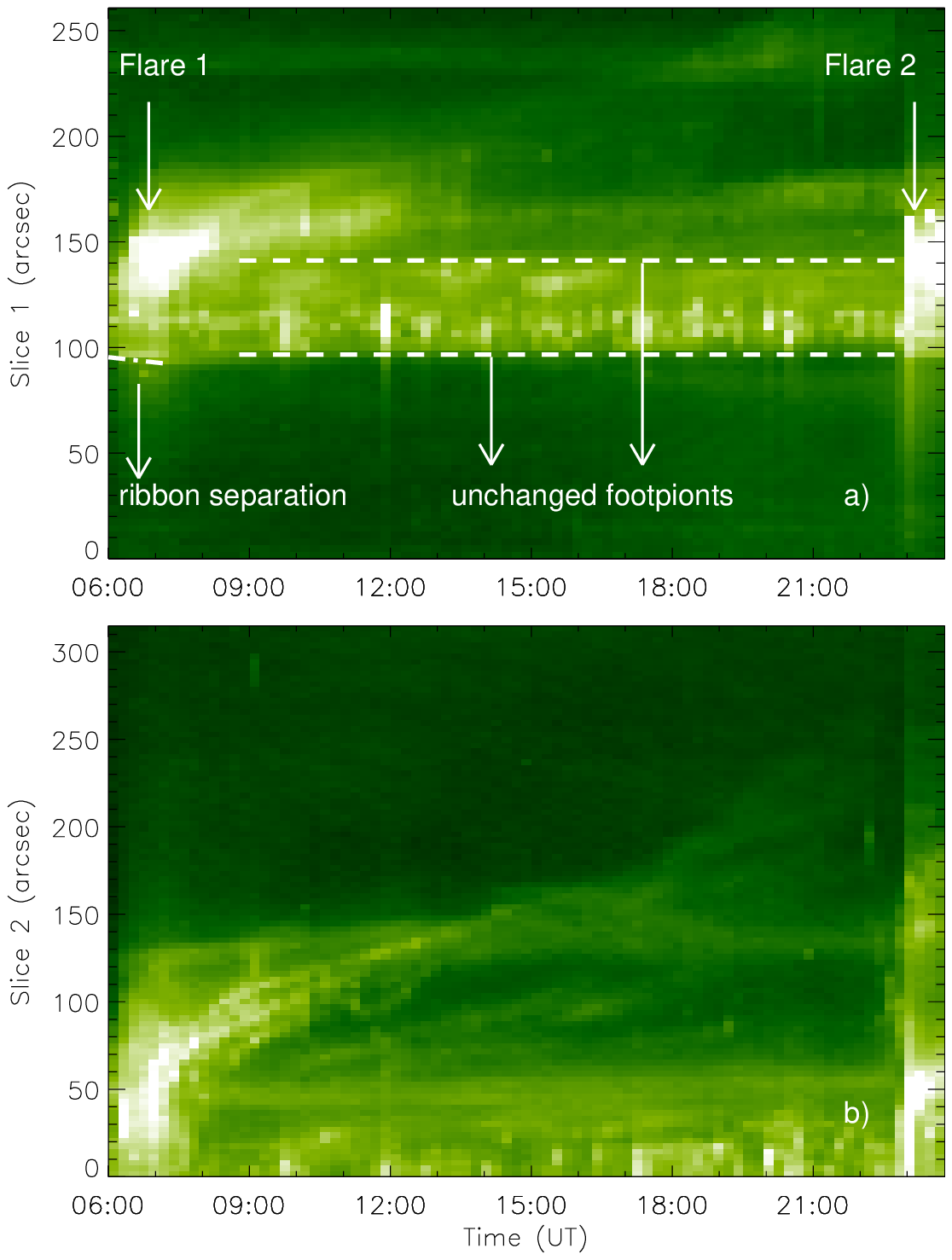} \epsscale{0.48}\plotone{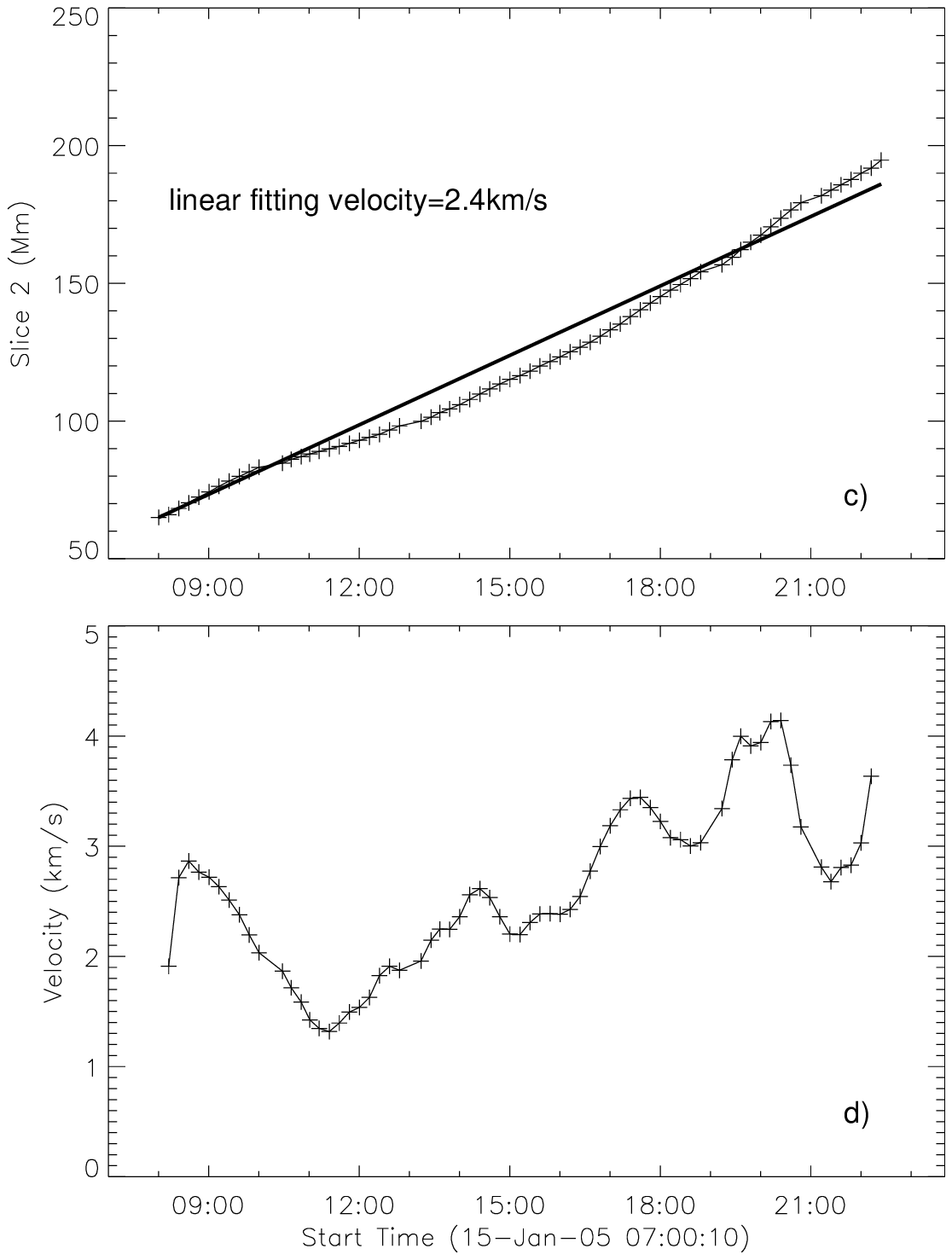}

\caption{(a--b) Time evolutions of brightness along slices 1 and 2
as shown in Figure 1f. The dash-dotted and dash lines in panel (a)
denote the separation of the southern flare ribbon and unchanged
footpoints, respectively. (c--d) Projected height and speed
evolution at the top position of the post-flare loops.
\label{slice}}
\end{centering}
\end{figure*}

In the standard flare model, ascending of the reconnection site in
the corona causes the apparent rising of the post-flare loops and
the flare ribbon separation \citep[e.g.,][]{kopp76}. In fact, this
is the manifestation of the successive stacking of reconnected loops
but not the real motion of the same loop. In this study, however, we
observed the real rising of a group of post-flare loops with their
footpoints keeping unchanged that were then activated as the second
CME. To identify that, we trace the progress of the post-flare loops
and their footpoints using two slices, one of which passes through
the projected top and the other crosses the footpoints of the
post-flare loops, as shown in Figure 1f. The left two panels of
Figure 2 show the time evolution of the brightness along slices 1
and 2, respectively. From the time evolution of slice 1, we find
that during the first flare, the southern flare ribbon had an
separation, as indicated by the dash-dotted line; about 1 hr after
the peak of the first flare, the footpoints of the post-flare loops
kept largely unchanged until the occurrence of flare 2, as indicated
by the two dash lines. Note that, the northern flare ribbon was
sheltered by the post-flare loops so that it was difficult to see.
Nevertheless, we can obtain a footpoint separation speed of 0.06 km
s$^{-1}$ if the uncertainty in the footpoint identification is
estimated to be 5.2\arcsec\,(2 pixels) during the period of $\sim$16
hr. Such a speed is far less than the typical separation speed of
flare ribbons of 10 km s$^{-1}$ \citep{qiu02,jing05}, which confirms
that the footpoints did not change. On the other hand, from the time
evolution of slice 2, one can notice that the top of the post-flare
loops was continually rising. We then measure the height-time
variation of the top position along slice 2, as shown in the right
upper panel of Figure 2. The linear fitting speed is 2.4 km
s$^{-1}$. The deprojected speed is $\sim$8 km s$^{-1}$ if we assume
that the loops rose radially, which is far more than the maximal
rising speed caused by the ascending reconnection site of $\sim$1.1
km s$^{-1}$ by \citet{hick85}. Furthermore, through the piecewise
numerical derivative method, the velocity evolution can be obtained
from the height-time measurement (Figure 2d). We find that the
rising of the post-flare loops was further accelerating, but not
decelerating, after 12:00 UT until the occurrence of flare 2. Based
on the standard flare model and the observations of
\citet{svestka96}, a rising motion of post-flare loops with a
decelerated or constant speed can be regarded as an apparent motion
resulting from the ascending reconnection site. In this letter,
however, we report a rising motion of the post-flare loops with an
accelerating speed that we think represents a real ascending motion
of the loops through the corona.

\begin{figure*}[!ht]
\begin{centering}
\epsscale{0.7} \plotone{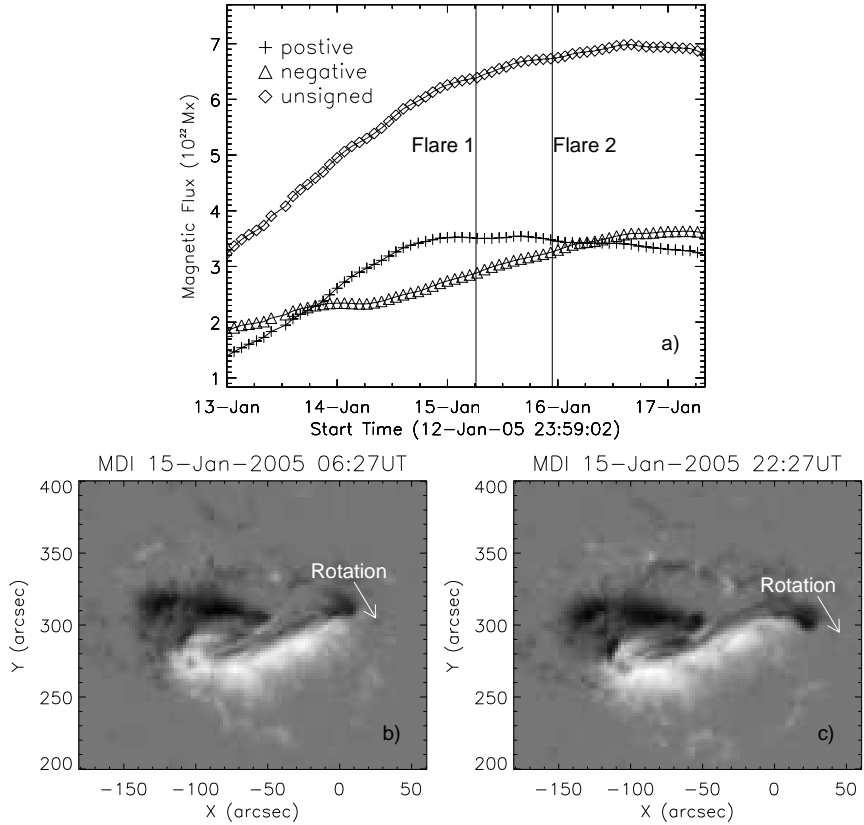} \caption{(a) Evolution of the
magnetic flux for the whole FOV of the active region. The two
vertical lines denote the onset times of the two flares. (b--c)
Line-of-sight magnetograms with a field of view of 240\arcsec
$\times$ 200\arcsec \ taken from MDI. The two arrows show the motion
direction of the western sunspot with a negative polarity.
\label{midflux}}
\end{centering}
\end{figure*}

We believe that the rising of the post-flare loops was driven by the
magnetic flux emergence and the rotation of the sunspot. The
evolution of the magnetic flux of the active region NOAA 10720 from
January 13 to 17 is plotted in Figure 3a. The two vertical lines
denote the onset time of flares 1 and 2. We note that the total
unsigned magnetic flux was continually increasing after the
occurrence of flare 1, which was mainly due to the rapid emerging of
the negative flux. We also plot two MDI line-of-sight magnetograms
around the onset time of the two flares in Figure 3b and 3c. After
comparing the two images, we find that the PIL of the bipolar active
region was elongated about 20\arcsec\,due to the westward motion and
rotation of the western sunspot with a negative polarity, as
indicated by the arrows in Figure 3b and 3c. Note that, the motion
and rotation of the negative polarity may be driven essentially by
the horizontal plasma flows in the photosphere. Usually, we know
that the magnetic free energy tends to decrease after the occurrence
of a major flare. Though, in some cases, one eruption does not
necessarily release the entire free energy of the active region, as
revealed in \citet{thalmann08}, who studied the time-series of a
NLFFF-equilibria computed from SOLIS. For the event in this study,
the continual magnetic flux emergence and the rotation of the
sunspot enhanced the magnetic pressure under the post-flare loops
and thus drove their rising and expanding motion; in the meantime,
they increased again the magnetic free energy in the corona. The
continually increasing magnetic free energy, in spite of the part
that was released during the first eruption, was still enough for
the second eruption of CME/flare.


\begin{figure*}[!ht]
\begin{centering}
\epsscale{0.9} \plotone{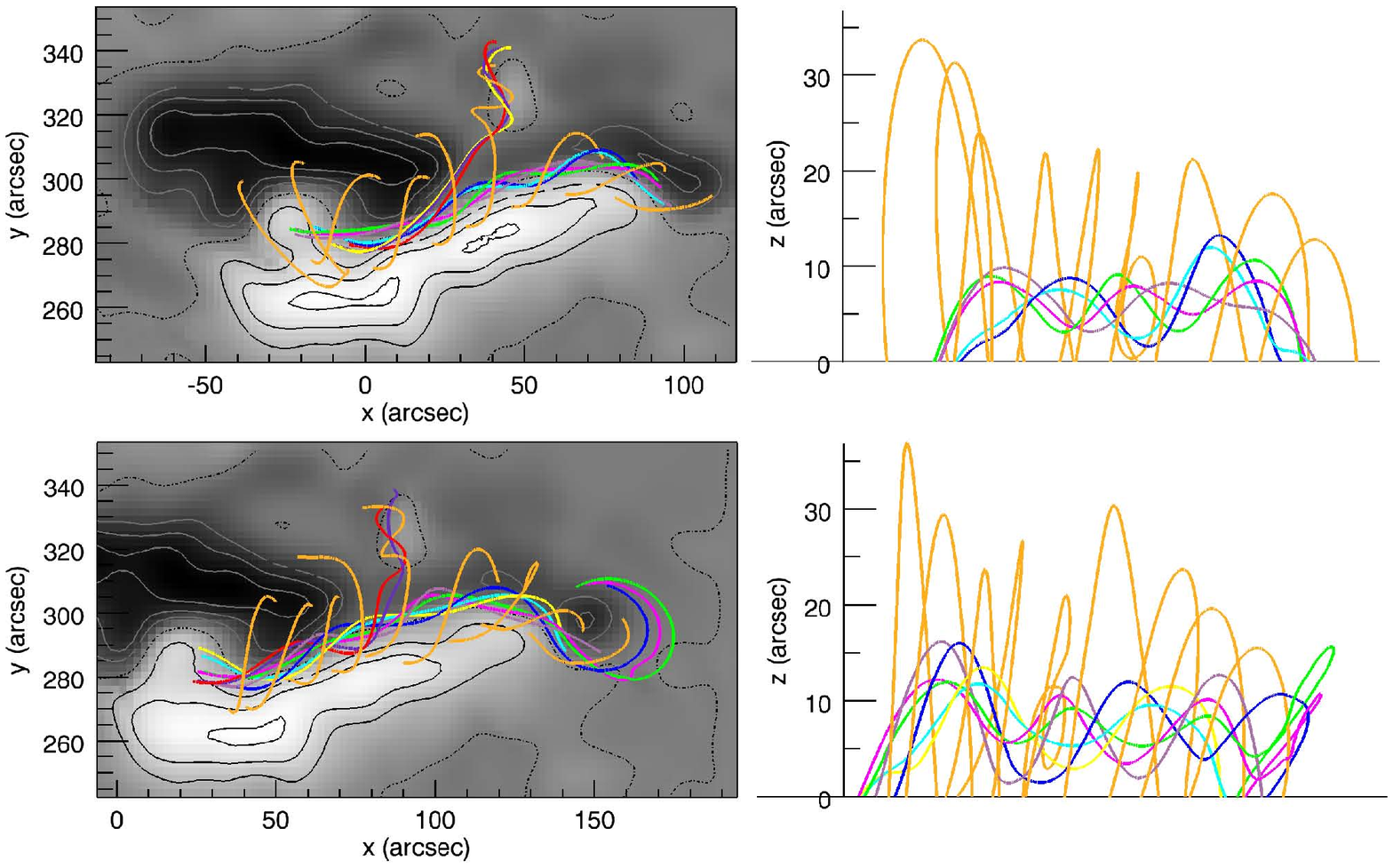}

\caption{Top view (a) and edge-on view (b) of the extrapolated 3D
magnetic field configuration at 18:00 UT on 2005 January 15. Panels
(c) and (d) are like (a) and (b), respectively, but for the time of
23:00 UT on 2005 January 15. In panels (a) and (c), the background
image is the processed line-of-sight magnetogram. Note that the
scale in $z$-direction is three times that in $x$- and
$y$-directions; the smaller flux rope in the northern branch in
panels (a) and (c) are not plotted in panels (b) and (d).
\label{fluxrope}}
\end{centering}
\end{figure*}

Another interesting finding is that the flux rope-like structure
remained to exist after the eruption of the first CME, which
continually ascended prior to the second CME. With the observed
vector magnetic field in the photosphere, we extrapolate the 3D
NLFFF configuration of this active region. The field lines are
plotted in Figure 4. The figure shows that the flux rope was under
significant development during the time period. It kept rising with
its length and twist increasing. In particular, the top of the flux
rope rose from $\sim$13\arcsec \ at 18:00 UT to $\sim$17\arcsec \ at
23:00 UT. The rising of the flux rope may provide a strong evidence
of flux rope emergence through the photosphere into the corona. In
addition, the developed flux rope may also have a role in enhancing
the magnetic pressure underneath the post-flare loop systems, which
tends to push the loop system upward. It is noteworthy that whether
the flux rope exists in CME/flare producing regions is still an open
and unresolved issue. Up to now, there are only a few studies that
showed the extrapolated flux rope based on the vector magnetic field
observation in the photosphere \citep[e.g.,][]{yan01,canou09,guo10}.
However, in this letter, we not only obtain a flux rope structure
with a strong twist but also its development. The detailed evolution
of the flux rope obtained by the DVMG vector field data with a high
cadence will be described in a future paper.

\section{Summary and Discussions}

In this study, we report the full evolution of a post-flare loop
system, characterized by the initial rising and expanding motion and
the final re-activation as another CME. We find that the rising and
expanding of the post-flare loops were essentially due to the
continual increase of the magnetic flux, which enhanced the magnetic
pressure underneath the post-flare loop system. On the other hand,
we speculate that the first CME/flare eruption may lead to a
weakening of the constraining tension force of the overlying field
to the post-flare loop system. We further study the development of
the associated flux rope by the NLFFF extrapolation, and find that
the twist and length of the flux rope were increased, which may be
due to the continual flux emergence and/or the rotation of the
sunspot. In fact, magnetic flux emergence \citep{fan01,fan03,fan04}
and sunspots rotation caused by photospheric plasma flows
\citep{amari00,torok03,torok05} are among the most plausible ways
that have been invoked in numerical simulations to increase the
magnetic non-potentiality leading to the final eruption.

Through a quasi-linear force free method \citep{wang01},
\citet{zhao08} studied the magnetic topology skeleton of this active
region. They found the rising of the magnetic null and its spine,
which is consistent with the rising of the extrapolated flux rope
and the strong twist of the flux rope as shown here. This strongly
twisted flux rope provides a source for the shear of the post-flare
loop system that enhances the non-potentiality of associated
magnetic configuration. \citet{jing09} studied the temporal
evolution of the magnetic free energy for this active region from
21:00 to 23:00 UT on January 15 and found that it tends to increase
prior to the onset of flare 2. The accumulation of the free energy
was mainly caused by the emergence of magnetic flux and the rotation
of the sunspots, which continually transfers energy below the
photosphere into the corona \citep{kurokawa02,cheng10b}. Therefore,
the post-flare configuration may still accumulate magnetic free
energy in a short time that is responsible for the second eruption.

It is worthy noting that the homologous flares can take place in the
same active region with similar morphologies and pattern of
development \citep[e.g.,][]{choe00}, and the successive flares
usually occur in the magnetic loops that are closely related but not
in exactly the same loops \citep[e.g.,][]{liu09}. Whereas, the most
significant difference, of the events in this studies from the
general homologous and successive flares, is that the second
CME/flare came directly from the post-flare loops of the first
eruption but not the loops structures in the similar or other
related locations.


In conclusion, it is commonly viewed that a post-flare loop system
evolves relatively uneventful after their formation following a
CME/flare eruption. Except for some dynamic flows along the
post-flare loops
\citep[e.g.,][]{1997SoPh..175..511B,1999SoPh..190..153Q}, the
magnetic structure of the loops is closer to potential than that
before the eruption, thus unable to re-erupt in a short time after a
major eruption. The event studied here provides an exceptional case
and may break such a point of view. It shows that a post-flare
magnetic configuration can evolve quickly to a state that is
favorable for another major CME/flare eruption, as long as there is
continual emergence of flux rope and/or rotation of sunspots.
Post-flare loops can re-flare in a short time. Models and prediction
of solar flares (CMEs) should take into account such an
observational fact.

\acknowledgements

We thank the anonymous referee for his/her valuable comments that
helped to improve the paper. X.C., M.D.D., and Y.G. were supported
by NSFC under grants 10673004, 10828306, and 10933003 and NKBRSF
under grant 2006CB806302. J.Z. was supported by NSF grant
ATM-0748003 and NASA grant NNG05GG19G. J.J. was supported by NSF
under grant ATM 09-36665 and ATM 07-16950. T.W. was supported by
DLR-grant 50 OC 0501. SOHO is a project of international cooperation
between ESA and NASA.

\bibliographystyle{apj}

\end{document}